\documentclass[12pt,preprint]{aastex}

\newcommand{\kms}{\mbox{\,km\,s$^{-1}$}}

\shorttitle{The optical light curves of XTE~J2123--058: paper III}
\shortauthors{Shahbaz et al., }

\begin{document}


\title{The optical light curves of XTE~J2123--058: 
the mass of the binary components and the structure of the
quiescent accretion disk}


\author{T.~Shahbaz\altaffilmark{1},
C.~Zurita\altaffilmark{1} 
J.~Casares\altaffilmark{1},
G.~Dubus\altaffilmark{2},
P.A.~Charles\altaffilmark{3},
R.~Mark Wagner\altaffilmark{4},
E.~Ryan\altaffilmark{5}}

\altaffiltext{1}{Instituto de Astrof\'\i{}sica de Canarias, 38200 La Laguna,
 	Tenerife, Spain}
\altaffiltext{2}{Theoretical Astrophysics, Caltech 130-33, CA 91125, Pasadena, USA }
\altaffiltext{3}{Department of Physics \& Astronomy, University of Southampton,
       Southampton, UK}
\altaffiltext{4}{Large Binocular Telescope Observatory, University of Arizona, Tucson,
       Arizona, USA}
\altaffiltext{5}{Department of Astronomy, University of Arizona, Tucson, Arizona, USA }



\begin{abstract}
We present optical photometry of XTE~J2123--058  during its quiescent state
taken in 1999 and 2000. The dominant feature of our R-band light curve is the
ellipsoidal modulation of the secondary star, however, in order to fit this
satisfactorily we require additional components which comprise an X-ray heated
Roche-lobe filling secondary star,  and an accretion disk bulge, i.e. where the
gas stream impacts the  accretion disk. The observed dip  near phase
0.8 is interpreted as the eclipse of inner parts of the accretion disk by the
bulge. 
This scenario is highly plausible given the high binary inclination. 
Our fits allow us to constrain the size of the quiescent accretion 
disk to lie in the range 0.26--0.56~$R_{\rm L1}$ (68 percent confidence). 
Using the distance of 9.6~kpc and the X-ray flux inferred from the  heated
hemisphere of the companion, we  obtain an unabsorbed X-ray luminosity of 
1.2$\times 10^{33}$\,erg\,s$^{-1}$ for XTE~J2123--058 in quiescence. 
From the observed quiescent optical/IR  colors we find that the power-law
index (-1.4) for the spectral  distribution of the accretion disk compares 
well with other quiescent  X-ray transients. 

We also re-analyse  the optical   light  curves of  the  soft  X-ray transient
XTE~J2123--058 taken during its  outburst and decay in 1998. We use a robust
method to fit the data using a refined X-ray binary model. The model  computes
the  light arising  from a Roche-lobe filling star and flared  accretion disk
irradiated by X-rays, and calculates  the effects of shadowing  and mutual
star/disk  eclipses. We obtain  relatively  accurate values   for  the 
 binary inclination
and mass ratio, which when combined with spectroscopic results obtained in
paper II  gives a neutron star mass in the range 1.04--1.56\,$M_{\odot}$  
(68\% confidence). 

\end{abstract}




\section{Introduction}

Soft X-ray transients (SXTs), or X-ray novae in quiescence provide outstanding
opportunities to study the mass donor and/or quiescent accretion disk in these
low-mass X-ray binaries (LMXBs) because of the absence of the enormous glare
that results from X-ray heated material during outburst.  The majority of these
systems appear to contain black-hole compact objects (see e.g. Charles 2001 and
references therein), but four of them have been identified with
neutron stars as a result of the type\textsc{i} X-ray bursts seen during X-ray
active phases.  The rarity of neutron star transients has been explained by
King, Kolb Burderi (1996) as due to the effects of X-ray irradiation which can
stabilize the accretion disk.

Furthermore, detailed studies of many transients are hampered by galactic
extinction, as most lie in or close to the galactic plane. And, as yet,
no transient has been seen to produce X-ray eclipses which would provide
incontrovertible constraints on its orbital inclination.  For these
reasons, the recent SXT XTE~J2123-058 is a system of considerable
importance in this field.  It is at high galactic latitude
($b_\textsc{ii}\sim$-36.2$^\circ$), and during outburst 
(Zurita et al., 2000; hereafter paper I) 
displayed a dramatic optical modulation (the largest
in this class of objects) which indicated that, even though not
eclipsing, the inclination must be very high. Observations of 
type\textsc{i} bursts at optical \citep{Tomsick98b} and X-ray wavelengths
\citep{Takeshima98} indicated that the compact object is a
neutron star, making it one of only four neutron star X-ray transients.

This paper is a companion to Casares et al., (2001; hereafter paper II)
which presents the first spectroscopic detection of the companion in
quiescence. Low resolution spectroscopy of XTE J2123-058 during its
quiescent state revealed a K7\textsc{v} star 
with a radial velocity 
semi-amplitude of 287$\pm$12~km~s$^{-1}$. Combined with the orbital 
period this implies a binary mass function of $f(M)$=0.61$\pm$0.08~M$_{\odot}$ 
(paper II). From these, estimates of the component masses were made, but 
it is clear that these are limited by the accuracy of the $q$
and $i$ values.

We present the optical light curve of XTE~J2123--058 taken when the source was
in quiescence.   We model the multi-component  optical light curves in terms of
the light arising from an X-ray heated Roche-lobe filling secondary star, the
accretion disk and the self eclipse of the inner part of the disk  by the
stream/disk impact region. We also take the opportunity to  re-analyse the
outburst and decay optical light curves of XTE~J2123--058. The model we used
previously suffered from computational biases, in particular when calculating
eclipses. Also, as is the case in most fitting procedures, the best fit
solution depends somewhat on the starting parameters. Here, we use a
genetic-type algorithm which is more robust than conventional techniques, which
allows us to determine accurate masses for the binary components. 

\section{Observations and data reduction}

Cousin R-band CCD images of XTE~J2123--058 in quiescence were obtained with the
NTT 3.5m telescope on La Silla, the Bok 2.3-m telescope at the Steward
Observatory, Arizona and with the WHT 4.2m telescope on La Palma between 1999
and 2000 (see Table~\ref{tble:log} for details). 
Johnson UBV and Cousin RI observations were also obtained
with the 4.2m WHT and INT 2.5m telescopes. In addition infrared (JHK)
observations were obtained during September 1999 as part of the UKIRT
Service Programme.

All the optical images were corrected for bias and flat-fielded in the
standard way. The infrared images were dark-current subtracted and then
flat-fielded using a median stack of the images. We applied
profile-fitting photometry to our object and several nearby comparison
stars within the field of view, using {\sc iraf}. All the images were
flux calibrated except for the J-band, because of the non-photometric
conditions during the J-band standard star observations.

\section{The colors of the accretion disk}

\begin{figure}
\plotone{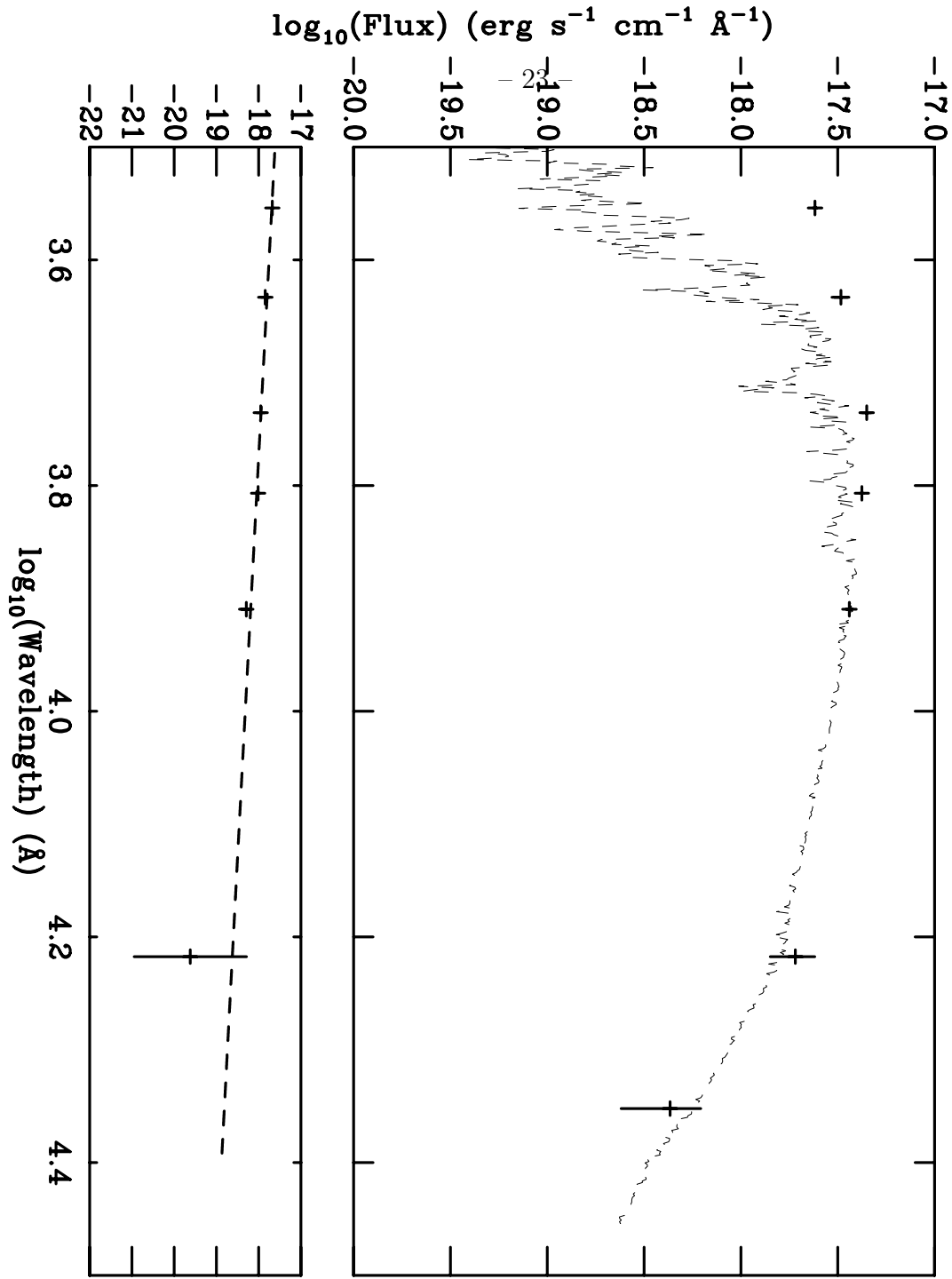}
\caption{Top panel: the dereddened UBVRIHK colors of XTE~J2123--058 are 
shown as crosses. The dotted line shows a $\tt PHEONIX$  model atmosphere 
spectrum of a K7\textsc{v} star scaled so that it contributes 77 percent to 
the R-band flux.
Bottom panel: the spectrum of the accretion disk obtained by subtracting the 
filter-folded K7 star colors from the dereddened colors. The spectrum has
a power-law shape with index -1.4.
\label{fig:colors} }
\end{figure}

We use the dereddened optical and infrared colors of XTE~J2123--058 to recover
the broad-band spectrum of the accretion disk by subtracting the companion
star. 
We use a $\tt PHOENIX$ model atmosphere  \citep{Hauschildt99} with 
$T_{\rm eff}$=4250 K and $\log g$=4.5 to represent the K7\textsc{v} 
secondary star, which is consistent with 
the spectral type classifaction by \citet{Tomsick01}.
The observed mean colors and dates are given in Table~\ref{tble:mags}. 
We  first convolved the model  atmosphere spectrum with the optical and infrared 
filter responses and  then fixed the secondary star to contribute 77 percent 
of the observed flux at $\sim$6300\AA\ (as determined in paper II). By 
subtracting the resulting colors from the observed dereddened colors of 
XTE~J2123--058 based on $E_{\rm B-V}=0.12$ \citep{Hynes01} we obtain the 
colors of the accretion disk (see Figure \ref{fig:colors}).  We find 
that the spectrum of the accretion disk can be represented by a power-law 
of the form  $F_{\lambda} \propto \lambda^{n}$ with index -1.4.

If we use $T_{\rm eff}$=4500~K to represent a K5 secondary star,
we obtain an  exponent of -1.1.
Changing the secondary star's contribution to 70 percent increases
the exponent by 0.3.
Therefore we estimate a systematic  uncertainty of $\pm$0.3.
Since the optical and infrared observations were taken  at different times we
have to assume that the mean brightness of the  source does not vary with time.
Ideally we would like to determine the phase averaged magnitudes. Although most
of our measurements do not  cover a complete orbital cycle we believe that the
large uncertainties in  the measurement for each band are such that we allow
for this systematic  effect in our error estimate.
The power-law index for the spectral
distribution of the accretion disk compares well with that derived in other
quiescent SXTs.   In A0620--00 and Cen X--4 the power-law index is
-2.5$\pm$1.0  \citep{McClintock86} and -1.6$\pm$0.30 \citep{Shahbaz93}
respectively.

\section{A re-analysis of the outburst/decay data}

In this section we re-fit the optical light curves of XTE~J2123--58 presented
in paper I.  The model that was originally 
used (paper I) suffered from many computational biases,
the result of many transfers of the code to different operating systems which
resulted in the miss-calculation  of eclipses. The errors in the code were
noticed when we tried to produce  simple images of the binary system. Also, the
code was not general and could only be used for systems with relatvely hot
secondary stars as is the case in high-mass X-ray binaries \citep{Tjemkes86}
and so it used Kurucz model atmospheres with temperatures
hotter than 5500~K. 
Therefore we took the opportunity to write and fully test a
new code that models the light from an X-ray irradiated Roche-lobe
filling star and accretion disk orbiting a compact object.  The model includes
a flared accretion disk, and new model atmosphere fluxes and limb-darkening
coefficients for cool stars.  Also, we now use a robust genetic-type 
algorithm \citep{Storn95}
to fit the light curves.  Unlike the algorithm used in paper I and most 
other studies, Genetic codes have been used before in fitting light curves, as
in \citet{Metcalfe99}, \citet{Gelino01} and \citet{Orosz02},
the  best fit solution does not depend on the starting parameters. 
Hence we are able to improve on the accuracy of the mass determination
for the  binary components of XTE~J2123--058.

\subsection{The X-ray binary model}

To interpret the optical light curves, we used a model that includes a
Roche-lobe filling secondary star, the effects of X-ray heating on the
secondary, a concave  accretion disk,  shadowing of  the secondary star by the
disk, and mutual eclipses of the  disk and secondary star .
For details of the model see \citet{Tjemkes86} and \citet{Orosz97}.
The geometry of
the binary system is given by the orbital inclination  $i$, the mass ratio 
($q=M_{2}/M_{1}$, where $M_1$ and $M_2$ are the  masses of the compact object
and secondary star respectively) and the  Roche-lobe filling factor $f$ of the
secondary star. The light from the  secondary star is  given by its mean
effective temperature  $\bar{T_{\rm eff}}$, the gravity  darkening 
exponent  $\beta$ and the X-ray albedo
$W$. The temperature  across the star is  scaled such that it weighted mean
matches the observed temperature.   
The  light  from the  accretion disk is given by its radius $R_{\rm disk}$,
defined as a fraction  of the  distance to the  inner Lagrangian  point
($R_{\rm L1}$),  its flaring angle, $\alpha$, the temperature  at the outer
edge of the disk  $T_{\rm out}$  and the exponent on the power-law radial
temperature  distribution $\xi$. Finally, the additional light due to X-ray 
heating  is given by the  unabsorbed  X-ray flux $F_{X,0}$,  the distance to 
the  source  $d_{\rm kpc}$ and the orbital separation (determined from  the 
optical mass function $f(M)$, $P_{\rm orb}$ the $q$ and $i$.). 
The X-ray heating of the secondary star is computed in the same way as 
described by \citet{Tjemkes86}.
We assume that
the secondary is in synchronous rotation and completely  fills its Roche lobe.
Since the late-type secondary star has a convective  envelope, we fix  the
gravity darkening  exponent to 0.08 \citep{Lucy67}.    
The albedo of the companion star ($W$) is fixed at 0.40  \citep{deJong96}. 
For ease, the key variables are listed in Table~\ref{tble:defs}.

\subsubsection{The accretion disk}

The  model  assumes a  flared,  concave  accretion  disk of  the  form
$h\propto  r^{9/7}$ with  $h$ and $r$ the disk height and radius respectively
\citep{Vrtilek90}. The radial distribution of the temperature  across the disk
is given by  $T(r)=T_{\rm out}r^{\xi}$. For a steady-state,  optically thick,
viscous accretion disk, the exponent of the power-law  is -3/4 
\citep{Pringle81}.  For  a disk heated by  a central source  the exponent is
-3/7 \citep{Vrtilek90}.  The temperature of the accretion  disk's rim is fixed
at  $T_{\rm out}$.  We assume that the  disk radiates  as  a blackbody. For  a 
given local temperature  in  the  accretion disk,  we  calculate  the 
blackbody intensity over  the wavelength  range of  the  filter and then
convolve it with the response of the filter.

\subsubsection{The intensity distribution}

The intensity  distribution on the secondary star  is calculated using
$\tt PHOENIX$ model atmospheres  colors that are publically available
on   the  World   Wide  Web.    $\tt  PHOENIX$   is   a  multi-purpose
state-of-the-art   stellar   atmosphere   code  that   can   calculate
atmospheres and  spectra of  stars all across  the Hertzsprung-Russell
diagram  \citep{Hauschildt99}.   The colors  that  are available
cover  a range  of temperatures  (1700~K--30000~K) and  gravity ($\log
g$=3.5--5.0)  and wavelength,  and  have been  already convolved  with
standard filter responses.  For this code  to be of general use, it is
important  that one can  calculate model  atmospheres for  cool stars.
In order to accurately calculate the  intensity at each grid point on the
star, which  depends on  the local temperature  and gravity, we  use a
2--D bicubic spine interpolation  method \citep{Press92}
using the 2--D color grid.

We use a quadratic limb-darkening law to correct the intensity from
the star and accretion disk. The coefficients are taken from 
Claret (1998) which were computed using $\tt PHOENIX$ model atmospheres.
These  calculations extend the range of effective temperatures
2000~K--50000~K and gravity $\log g$ = 3.5--5.0.   Again we use 
a 2--D bicubic spline interpolation method to calculate the
coefficients for a given temperature and gravity.

\subsection{The fitting algorithm}

The minimization of  a function is a difficult  one, especially if the
function is  complicated and has  many parameters that  are correlated
and  even more  so if  there are  many isolated  local minima  or more
complicated  topologies.  Methods such  as Powell  (a maximum-gradient
technique),  downhill simplex (polyhedral  search technique)  are only
good if the global minimum lies near the initial guess values; this is
not normally  the case for  complicated functions. When  the objective
function is nonlinear, direct  search methods using algorithms such as
by  Nelder \& Mead (1965)  or genetic  algorithms  are  the  best
approaches \citep{Press92}.

The basic strategy at the core of every direct  search method  is to generate
variations of the parameter vectors and make a decision whether or not to 
accept the newly derived  parameters.  A new parameter vector  is accepted if
and  only if it reduces  the value of the objective function.  Although this
process converges fairly fast, it runs the risk  of becoming trapped by a local
minimum. Inherently  parallel search techniques like genetic and evolutionary
algorithms have some built-in safeguards to ensure convergence.   By  running
several vectors  simultaneously, strong parameter  configurations can help
other vectors escape local minima.

Ideally  an optimization  technique   should find the true global minimum,
regardless   of  the   initial  system   parameter   values and  should be 
fast. Storn \& Price (1995)  have  developed  a  genetic-type  technique 
called ``Differential  Evolution'' ,  which is  robust  and simple.   
Differential  Evolution generates  new parameter vectors by adding  the
weighted difference vector  between two  population  members  to a  third 
member. If  the resulting  vector  yields a  lower  objective  function  value
than  a predetermined population  member, the newly  generated vector replaces
the vector  with which it was  compared, in the  next generation.  The best
parameter vector is also  evaluated for every generation in order to keep 
track of  the progress that  is made during  the minimization process. There
are four  main parameters in the differential evolution code; $\tt  NP$, the 
number of population  members (usually  taken as 10$\times$  the number of 
fitting parameters);  $\tt F$  the mutation scaling  factor which  controls 
the amplification  of the  difference vector; $\tt CR$ the crossover
probability constant and  $\tt itermax$, the maximum number of generations. For
our problem of  minimizing the $\chi^{2}$ function we use $\tt CR$=0.9, $\tt 
FX$=0.9  and  $\tt itermax$=2000, which corresponds to $\sim$100,000
calculations  of the function.

\subsection{Fitting the outburst/decay data}

We use the differential evolution algorithm described in section 4.2
to fit the same phase-binned outburst and decay light curves of XTE~J2123--058 
which were presented in paper I.
We fix $d_{\rm kpc}$=9.6, $P_{\rm orb}$=0.2482~d and 
$\bar{T_{\rm eff}}$=4250~K 
corresponding to a K7 star (paper II). The observed X-ray flux
(ASM 2--12 keV energy range) during the time of the outburst and decay
data sets was 1.7$\times~10^{-9}$~erg~cm$^{-2}$~s$^{-1}$ and
1.7$\times~10^{-10}$~erg~cm$^{-2}$~s$^{-1}$ respectively.  The R-band
reddening is fixed to be 0.27 mags based on $E_{\rm B-V}=0.12$ 
\citep{Hynes01}. The model parameters that we fit are $q$, $i$, 
$R_{\rm disk}$,  $\alpha$, $T_{\rm out}$ and $\xi$.

\begin{figure}
\plotone{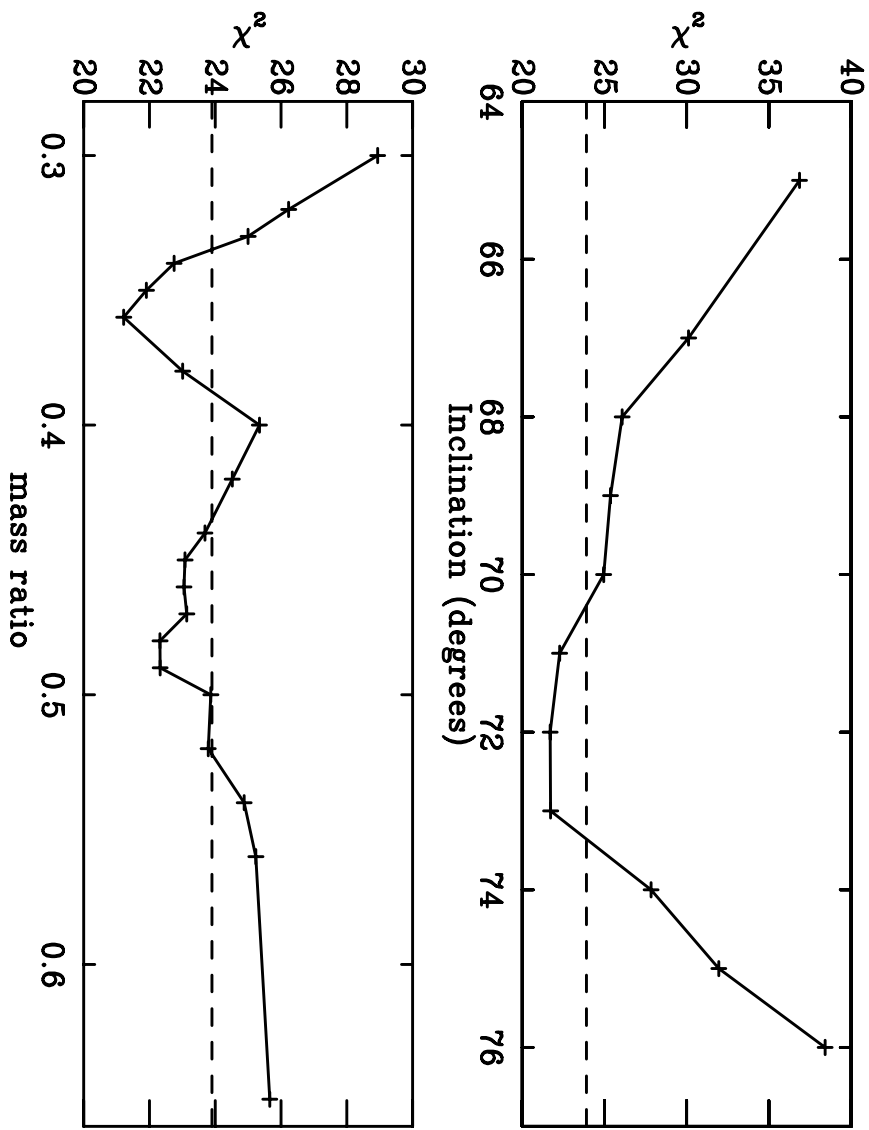}
\caption{The $\chi^{2}$ values obtained for the outburst data  by grid searching the
binary inclination  (top panel) and mass ratio (bottom panel)  whilst 
minimising the  other  model parameters.    The dashed line shows the 
90 percent confidence level.
\label{fig:chio} }
\end{figure}

For the outburst light curve with 29 data points
we find a minimum reduced 
$\chi^{2}$, $\chi^{2}_{\nu}$ of 0.9 at $q=0.36$, $i$=72.5$^{\circ}$, 
$R_{\rm disk}=0.69\,~R_{\rm L1}$, $\alpha$=5.9$^{\circ}$, $\xi=-1.26$ 
and T$_{\rm out}=2.5\times 10^4$\,~K.  In order to determine the 
uncertainty in the parameters of interest we perform a 1--D 
grid search and plot the 68 percent and 90 percent confidence levels 
(see Table~\ref{tble:fitod} and Figure~\ref{fig:chio}).  
If we use a distance of $d_{\rm kpc}$=11 the minimum $\chi^{2}$ is not
significantly different (at the 90 percent level). The inclination is the same,
however, $\alpha$ increases by 3.5$^{\circ}$. This is becasue the  model needs
to  increase $\alpha$  in order to shield the secondary star from the strong
X-ray heating 
(L$_{\rm X}$=2.5$\times 10^{33}$\,erg\,s$^{-1}$ for $d_{\rm kpc}$=11.0
compared with 
L$_{\rm X}$=1.9$\times 10^{33}$\,erg\,s$^{-1}$ for $d_{\rm kpc}$=9.6).

\begin{figure}
\plotone{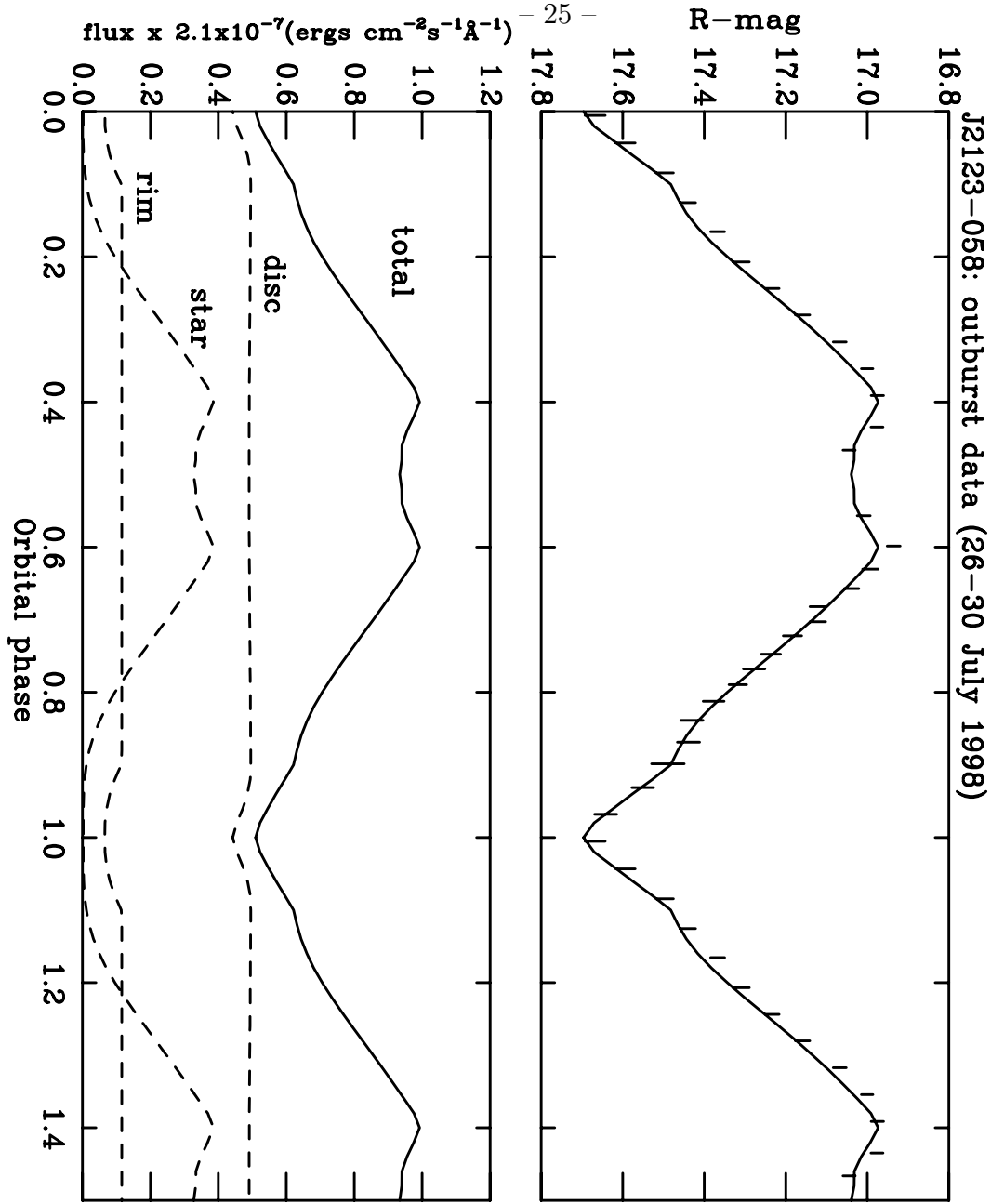}
\caption{Top  panel: the  average R-band orbital light curve during 
outburst (26--30  July 1998) of 
XTE~J2123--058 (vertical line) and the best model fit (solid line). 
The vertical lines represent the error in the mean of each phase bin.
Bottom panel: the three components  in  the X-ray binary  model  which gives  
the  best  fit to the outburst light curve.  The accretion  disk and rim  
are shown along  with the X-ray heated secondary star and the total 
flux. For clarity 1.5 orbital cycles are shown.
\label{fig:fito}  }
\end{figure}

\begin{figure}
\plotone{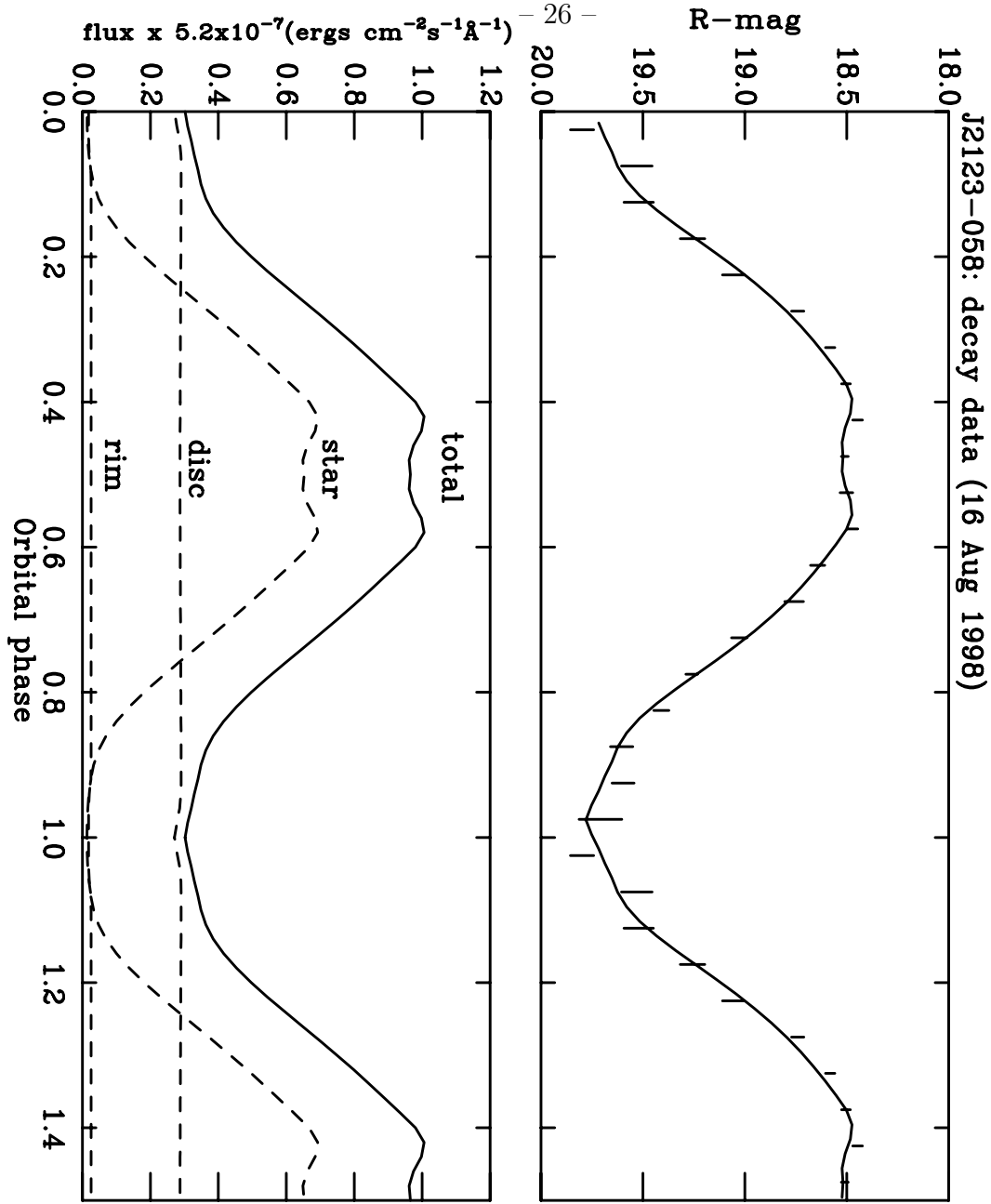}
\caption{Top panel:  the  average R-band orbital light curve during  
decay (16  Aug 1998) of  XTE~J2123--058 
(vertical line) and  the  best  model  fit  (solid  line).  
The vertical lines represent the error in the mean of each phase bin.
Bottom  panel:  the  three components  in  the X-ray binary model  which gives  
the  best  fit to the outburst data.  The accretion  disk and rim  
are shown along  with the X-ray heated secondary star and the observed 
flux. For clarity 1.5 orbital cycles are shown.
\label{fig:fitd}  }
\end{figure}

Since the quality of the outburst light curve is much better than the decay
light curve, the  eclipse features  at orbital phases 0.0 and 0.5 are more 
evident. Hence, the parameters derived from modelling the outburst light curve
will be more accurate compared to those derived from the decay light curve.
Therefore for the decay light curve with 20 data points we fix $i$ and $q$
derived from the  outburst light curve, and determine $R_{\rm disk}$ and
$\alpha$ of the  accretion disk.  
We find a minimum $\chi^{2}_{\nu }$ of 1.5 at 
$R_{\rm disk}=0.56\,~R_{\rm L1}$, $\alpha$=5.2$^{\circ}$, $\xi=-1.19$ 
and T$_{\rm out}$=7505\,~K (see Table~\ref{tble:fitod}).  The best model 
fits to the outburst and decay light curves are shown in 
Figures~\ref{fig:fito} and ~\ref{fig:fitd} respectively.
We also show the different components of light in the model, i.e. the 
accretion disk, the rim of the disk and the X-ray heated secondary star.
It should be noted that changing $\bar{T_{\rm eff}}$ does not have 
a significant effect on the light curves; its effect is only significant at the 
0.4 percent level. This is because of the dominant effect of X-ray heating.
Changing the albedo of the companion star to 0.30 and 0.5 
(de Jong, van Paradijs \& Augusteijn 1996; Milgrom \& Salpeter 1975) gives a
worse $\chi^2$, but it is only significantly different at the 95 percent level.
The derived parameters are the same within the uncertainties.

We can compare the results we obtained from fitting the outburst and decay 
light curves with our new model and fitting procedure with the those
obtained in paper I. The outburst values obtained for $i$ and $\alpha$ 
in paper I were significantly
overestimated (99\% confidence level) compared to those we obtain here.
Also, the value we obtain for $q$ in  paper I was
significantly  underestimated ($>$99\% level) compared  to the value 
given here. 
The results obtained through fitting the decay light curve  agree well
with those obtained in paper I.   The difference in the results 
compared to paper I is due mostly to the  errors in
the code and also in the fitting method  used
to find  the global minimum solution (see section 4).  
The most likely reason why the results from the decay light curve are 
comparable is the relatively  large error bars of the decay light curve. 
This would
naturally give  larger uncertainties in the derived parameters and so it is not
surprising that the results obtained using the model in paper I and our new
model agree.

\citet{Vrtilek90} showed that the temperature profile of a flared
accretion disk totally dominated by irradiation 
(i.e. isothermal) takes the form $T(r) \propto r^{-3/7}$ rather
than $T(r) \propto r^{-3/4}$ for a steady-state non-irradiated
disk. The value for the temperature exponent we derive is more negative
($\sim -1.2$) than either of these values.  For a fixed outer disk
temperature (and geometry)  a steeper exponent implies a hotter disk.
Therefore in order to match the observed flux using $\xi=-3/7$, the
projected area of the disk in the outer parts must be on the whole
smaller. Note that this can be achieved if the disk's structure is
warped; it has been known for some time that X-ray binaries can produce
warped accretion disks (Wijers \& Pringle 1999; Ogilvie \& Dubus 2001).    
In order to explain the observed reprocessed X-ray 
flux in LMXBs, the disk in these systems must be either warped or 
the central X-ray source is not point like \citep{Dubus99}. 
With this dataset we cannot
determine the shape of the accretion disk. However, it should be noted that
multi-color observations throughout an outburst may allow one to
determine $\xi$ (see Orosz \& Bailyn 1997) and more 
importantly provide information
on the shape of the accretion disk in an SXT during outburst.

\subsection{The shape of the light curves}

At the start of an outburst angular momentum in the disk is transferred to the
outer parts and so the matter diffuses inwards and the disk radius expands.
When the system's brightness decays,   the disk shrinks in radius.  The changes
observed in the light curves  during decline are primarily caused by large
changes in the disk size  and geometry.   As we can see from the different
component fits to the outburst and decay data, in outburst, the irradiated disk
light swamps the light from the irradiated secondary star and hence the
amplitude of the light curve  is relatively small.   As the irradiation of the
disk becomes less the disk temperature  is much cooler and the disk radius
smaller,  so the fraction of light from the disk is small compared to the
fraction of light from the secondary star.  This results in a large amplitude
for the light curve.  The triangular shaped minimum at phase 0.0 in the
outburst light curve can be interpreted as the eclipse of the accretion disk by
the X-ray heated secondary star. The deep mimimum in the outburst data  at
phase 0.5 i.e.  when the X-ray heated secondary star is eclipsed by the disk,
implies a larger disk radius compared to the disk radius when the X-ray
irradiation is less.
The same is true for the eclipse features at phase 0.0.   The fit to the decay
data requires a smaller and cooler, but not significantly thinner, accretion
disk compared to the disk in the outburst data.  Our model fits imply a change
of $\sim$20 percent in the disk size.

\section{The binary masses}

\begin{figure}
\plotone{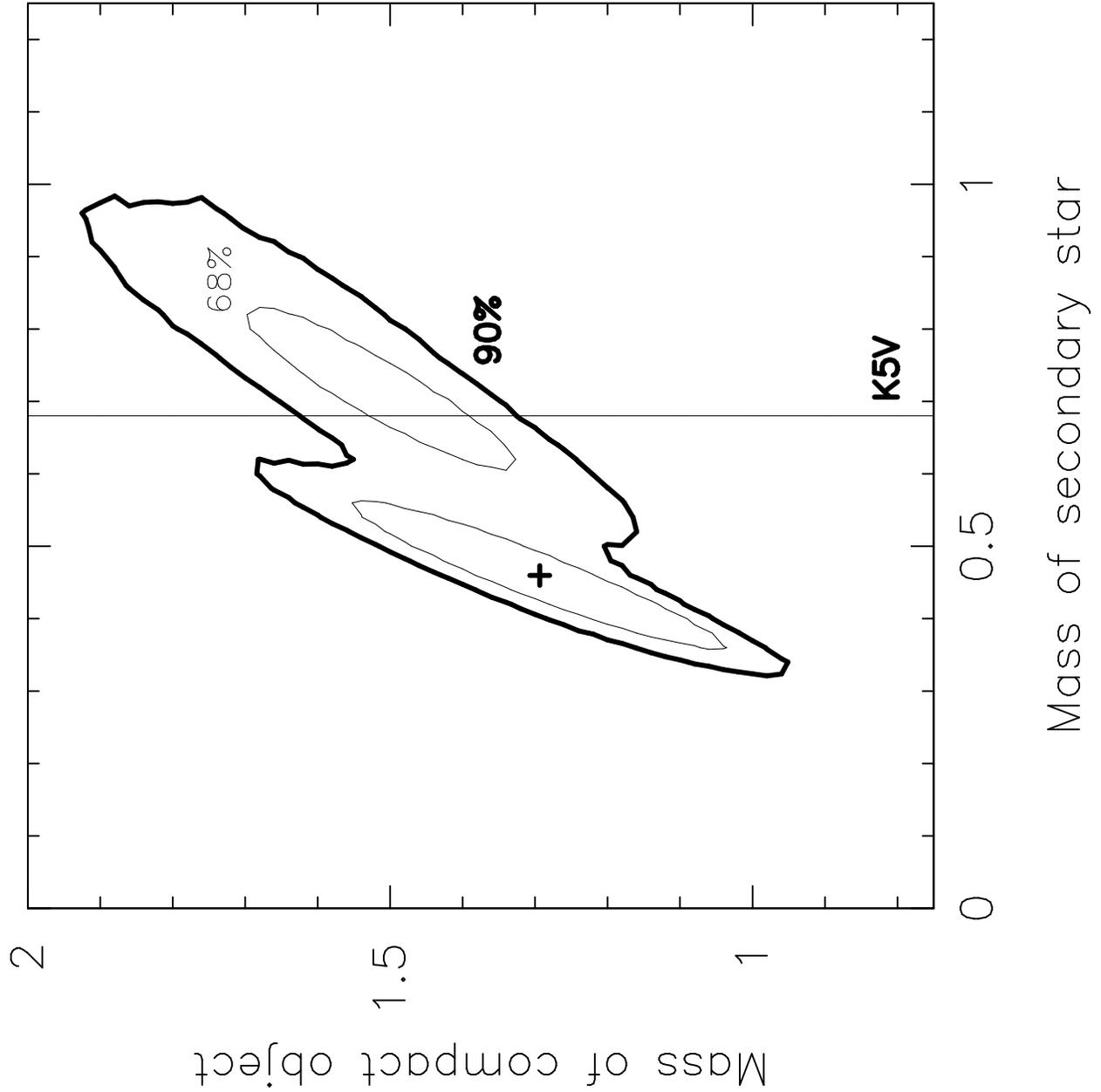}
\caption{The  mass  of the  binary  components  obtained  using a  
Monte  Carlo simulation  of the  observed parameters  and their  
uncertainties (see section 5). The  68 percent and 90 percent confidence 
levels  are marked as the thin and thick lines respectively. The vertical line
shows the mass of a ZAMS K5V star.
\label{fig:mcarlo}}
\end{figure}

The masses of  the binary components are given by  re-arranging the 
equation for the optical mass function,

\begin{equation}
M_{1} = \frac{K^{3}_{2} P_{\rm orb}}{2 \pi G}
\frac{(1+q)^2}{\sin^{3}i}~~M_{\odot};~~~~~~
M_{2} = q M_{1}, 
\end{equation}

\noindent 
where $K_2$ is the radial velocity semi-amplitude and $G$ is the 
Gravitational constant. Substituting our values for $q$ and $i$ with 
$K_{2}$ and $P_{\rm orb}$ we calculate $M_{\rm 1}$ and $M_{\rm 2}$.  
In order to determine the uncertainties we use a Monte
Carlo simulation, in which we draw random values for the observed
quantities which follow a given distribution, with mean and variance the
same as the observed values. For $K_{2}$ and $P_{\rm orb}$ the random
distribution are taken to be Gaussian because the uncertainties are
symmetric about the mean value;  $K_2$=287$\pm$12\kms, 
$P_{\rm orb}$=0.248236$\pm$0.000002 days(1-$\sigma$ errors; paper II). 
However, for $q$ and $i$ the uncertainties are asymmetric and so we 
determine the actual distribution numerically.  This is done by first 
calculating the maximum likelihood distribution using the actual 
$\chi^{2}$ fit values (see section 4.2)  and then determining the 
cumulative probability distribution. By picking random values (from a 
uniform distribution) for the probability, we obtain random values for 
$q$ and $i$.  Figure~\ref{fig:mcarlo} shows the results of the Monte Carlo 
with 10$^{6}$ simulations. Table~\ref{tble:mcarlo} gives the ranges we 
obtain for the system parameters.

\citet{Tomsick01} constrain the spectral type of the secondary star to lie
in the range K5--K9.  We can use the earliest spectral type (K5) for the
secondary star to place an upper limit to the mass of the secondary star, since
the mass of the secondary  must be less than or equal to the main sequence mass
of the  same spectral type. 
A K5V  star has a  mass of ZAMS mass of 0.68\,$M_{\odot}$ \citep{Gray92}.  
Although the present-day mass of the secondary star is lower, and the
secondary star is metal poor and hence will have a lower mass compared to a
normal star,  we can still use the ZAMS mass as a firm upper limit to the mass
of the secondary star. 
Using this limit we constrain   $M_{\rm 1}$ to lie in the range
1.04--1.56\,$M_{\odot}$ and 0.95--1.68\,$M_{\odot}$ 
(68\% and 90\% confidence respectively), which is
consistent with the mass range
determined  in paper II and by \citet{Tomsick01}.

\section{The quiescent light curve of XTE~J2123--058}

\begin{figure}
\plotone{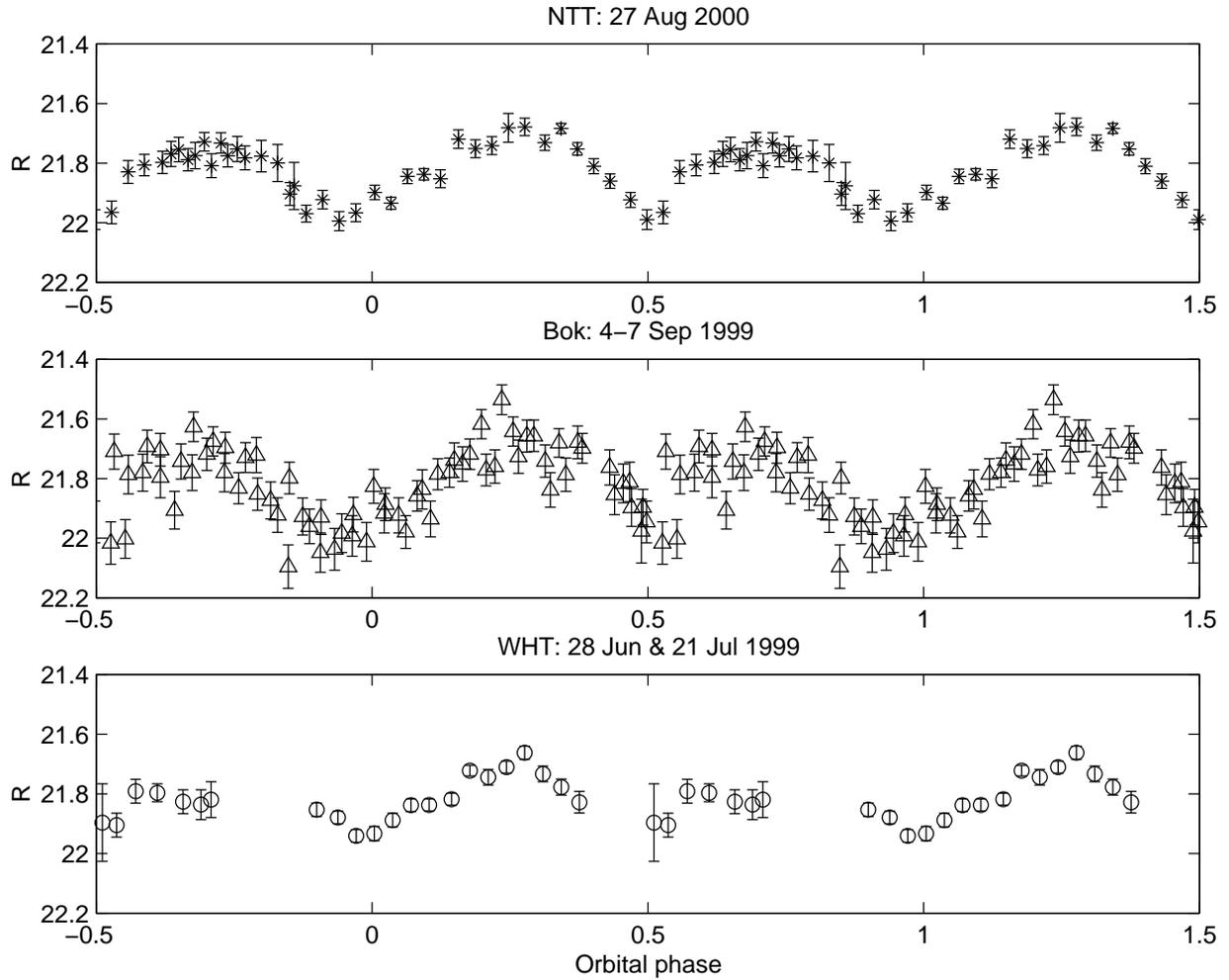}
\caption{The observed quiescent R-band light curves of XTE~J2123--058. 
From top to bottom: the NTT (August 2000), Bok (September 1999) and 
WHT (June/July 1999) data respectively. Note the stability of the light 
curves from epoch to epoch. For clarity 1.5 orbital cycles are shown.
\label{fig:quietlc} }
\end{figure}

In Figure~\ref{fig:quietlc} we show our new quiescent optical light curve of
XTE~J2123--058 obtained at different times, phase folded using the ephemeris
given in paper II.  Overall, the shape and mean brightness level of the light
curves are very similar for data taken at three different epochs over a year.
This is in contrast to the optical light curve of other SXTs, e.g. Cen X--4 and
A0620--00 whose light curves are known to change shape and mean brightness 
(McClintock \& Remillard 1990; Leibowitz, Hemar \& Orio 1998;  Gelino,
Harrison \& Orosz 2001). Such changes have been attributed to dark spots on the
surface of the late-type companion star, which are thought to migrate on
timescales of a few months to years \citep{Bouvier98}.

\begin{figure}
\plotone{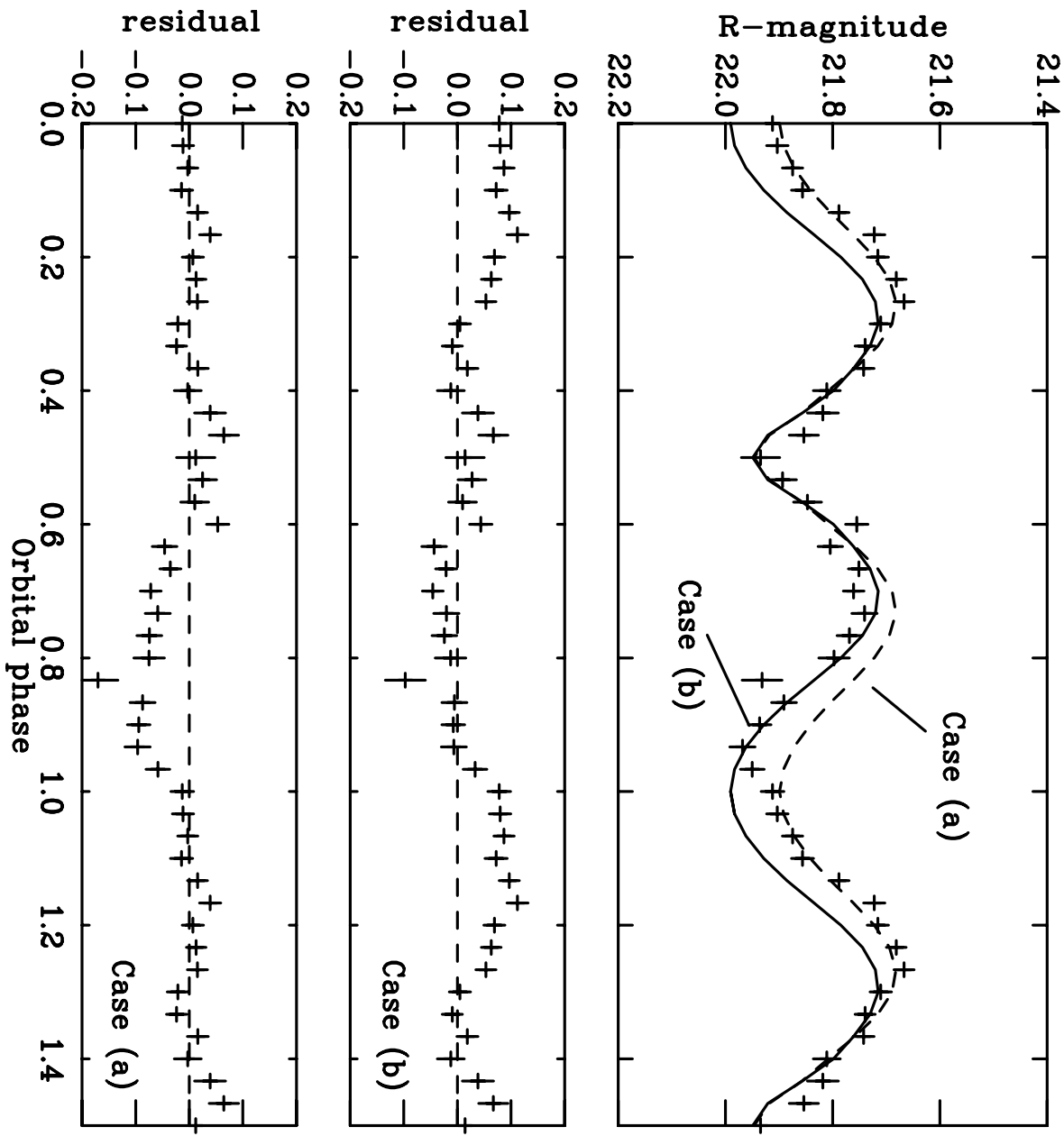}
\caption{Top panel: Model fits for case (a) and  case (b) (see text):
middle and bottom panels are the residuals for case (b) and (a) respectively.
\label{fig:caseab} }
\end{figure}

One expects the quiescent light curves of X-ray transients to be dominated
by the ellipsoidal variations of the secondary star, which gives rise to
a double-humped modulation, where the maxima are equal and the minima are
unequal for high inclination systems. However, the quiescent light curve
of XTE~J2123--058 is far from being clean;  the minumum at phase 0.0 
(inferior conjunction of the secondary star) appears to be offset by
$\sim$0.04 in phase.  In an attempt to try and understand this light
curve we consider two cases: (a) where we match the amplitude of the
light curve between phase 0.0 and 0.5 and (b) where we match the
light curve between phase 0.5 and 1.0 with our X-ray binary model (see
Figure~\ref{fig:caseab}).

From the residuals for these two cases we can see that there is either
excess light around phase 0.15 [case (a)] or there is a lack of
light i.e. a dip around phase 0.8 [case (b)].  The excess light near phase 0.15 does
not coincide with where a bulge is expected i.e the region where the
gas-stream impacts the edge of the accretion disk;  bright spots normally
precede the secondary star and are observed around orbital phase
$\sim$0.8--0.9 [see Wood et al., (1986) and Wood et al., (1989) for 
examples of the bright spot in high mass ratio quiescent dwarf novae]. 
Given the high binary
inclination of XTE~J2123--058 one can envisage situations where the inner
part of the accretion disk is eclipsed by a bulge.
If the bulge itself
does not emit much light compared to the accretion disk, then the net
effect on the light curves would be a decrease in light near phase 0.8.
Similar effects have been seen in the 
UV light curves of Z~Cha during superoutburst. In Z~Cha the dip at 
phase 0.8 was explained as being due to vertical disk structure occulting 
the hot innner disk \citep{Har92}.
Therefore, in the next section we assume that the dip near
phase 0.8 is due to the bulge eclipsing the inner parts of the accretion
disk.

\subsection{Fitting the quiescent light curve}

The three quiescent R-band light curves of XTE~J2123--058 were combined and
binned into 30 phase bins. The parameters in the X-ray binary model that we
vary are $i$, $\bar{T_{\rm eff}}$, $F_{\rm X,0}$, $R_{\rm disk}$, $T_{\rm out}$
and $\xi$.  We fix $d_{\rm kpc}$=9.6 and the R-band reddening 
to  be 0.27 mags.
In our initial fits we let $\alpha$ run as a free parameter but we found  that
it did not make a significant difference to the fits (in fact an F-test on the
best fit values does not justify the addition of an extra parameter), therefore
we fixed its value to 2$^{\circ}$, i.e. we assume a geometrically thin
accretion disk. Given the determination of the veiling of the R-band light, we
impose the condition that the accretion disk cannot contribute more than 33
percent to the observed light (using the upper limit of 77\% for the secondary
star's contribution to the obseved flux; see paper II).   When one determines
the veiling of an absorption line spectrum, one compares the strength of the
absorption lines of a template star (which normally has solar metalicity) with
the target spectrum. If the target has a metal-poor spectrum, then the veiling
will be overestimated. Given the high galactic latitude and systemic velocity
of XTE~J2123--058, the possibility that it is a halo metal poor system cannot
be ruled out. Hence we use the veiling estimated in paper II as a firm {\it
upper} limit to allow for the possibility that the veiling may be
overestimated. 
We do not attempt to include an accretion disk bulge in the X-ray binary  model
because  the height and temperature of the bulge most probably  varies with
orbital phase, thus introducing many parameters that would  be difficult to
deconvolve. Instead we model the accretion disk light  eclipsed by the bulge as
a gaussian function  which is included in the fitting algorithm.

\begin{figure}
\plotone{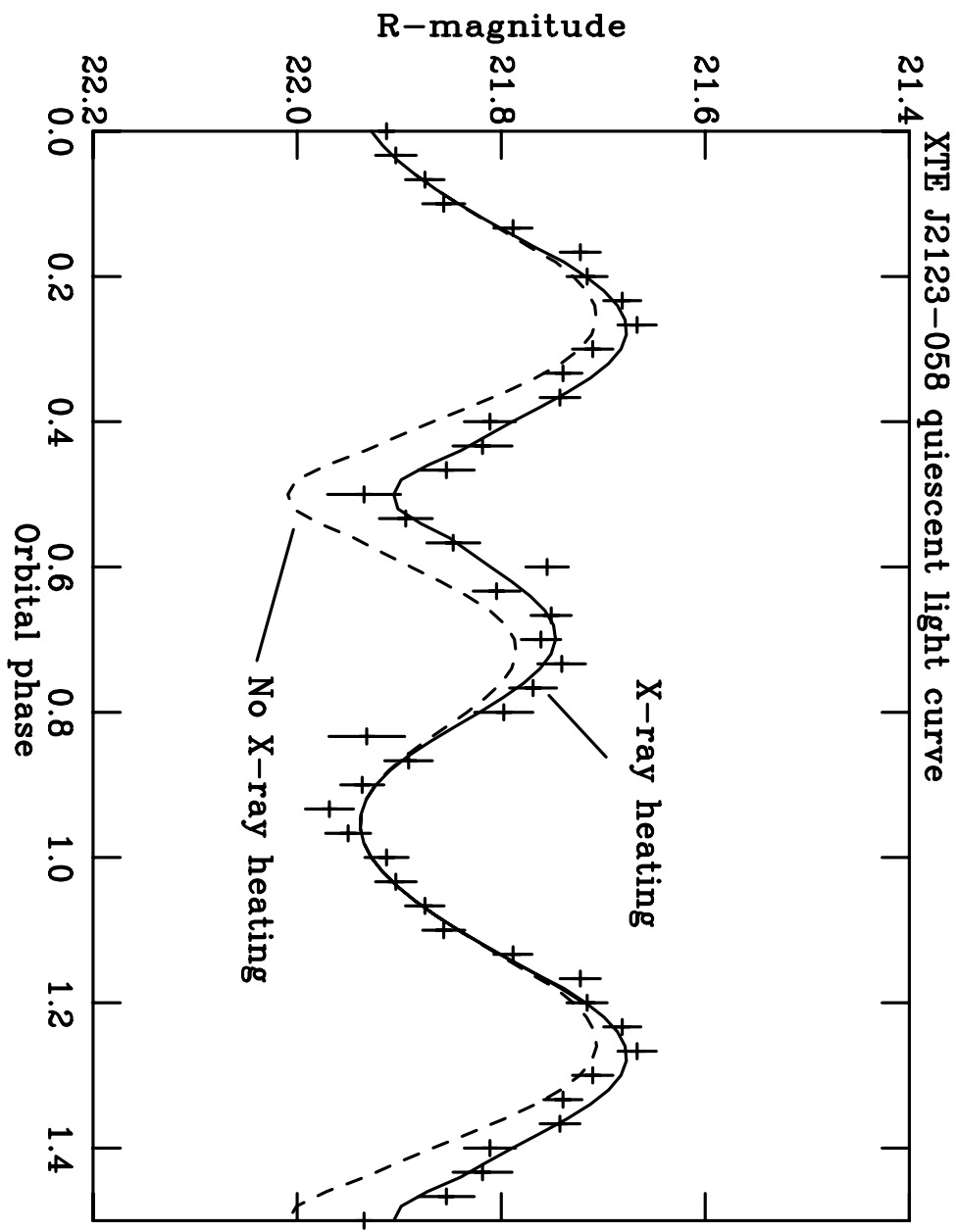}
\caption{The  observed quiescent R-band light curve of XTE~J2123--058
(crosses) and the best model fit (sold line). The dashed lines show 
a model with no X-ray heating. For clarity 1.5 orbital cycles are show.
\label{fig:fitq} }
\end{figure}

\begin{figure}
\plotone{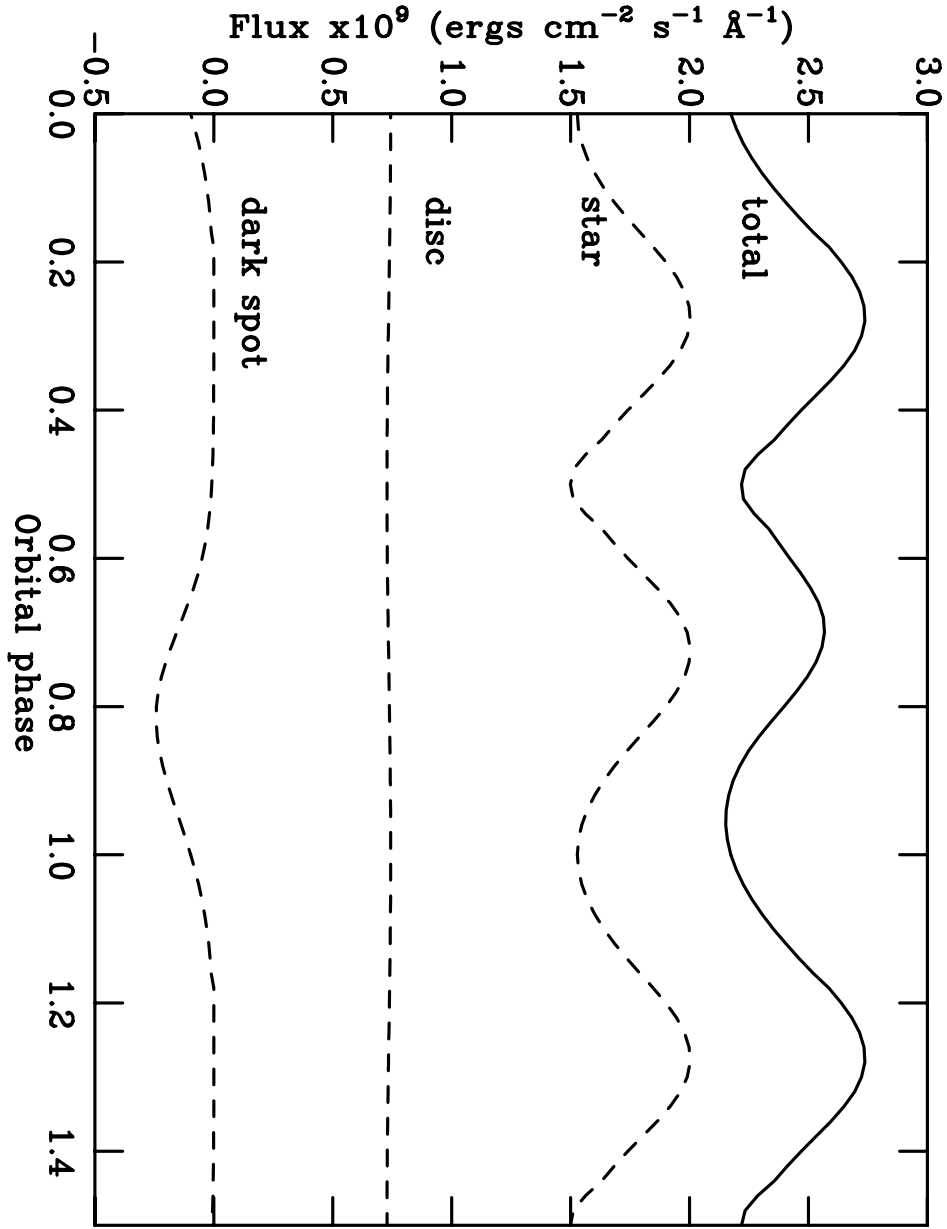}
\caption{The separate components  in the X-ray  binary model  which gives 
the best fit to the quiescent data. The accretion disk and bulge are shown  
along  with the X-ray heated secondary star and the total observed flux. 
For clarity 1.5 orbital cycles and dereddened fluxes are shown.
\label{fig:fitqc} }
\end{figure}

We use the differential evolution algorithm (see section 4.2) to fit the
phase-binned light curve. 
For the fits with $q$=0.36, we find a minimum $\chi^{2}_{\nu}$ of 1.2 at
$i$=71.8$^{\circ}$, $\bar{T_{\rm eff}}=4610$~K, T$_{\rm disk}=1425$~K, 
$R_{\rm disk}=0.45\,~R_{\rm L1}$, 
$\log F_{\rm X,0}$=-12.99~erg~cm$^{-2}$~s$^{-1}$ and $\xi=-1.23$.  
The bulge is centered at 0.84$\pm$0.01 phase and has a gaussian width and 
height of 0.12$\pm$0.01 phase and -0.12$\pm$0.01 mags respectively. 
A 1--D grid search was performed to determine the uncertainties in the fitted
parameters (see Table~\ref{tble:fitq}). 
For the fits with $q$=0.48, the minimum $\chi^{2}$ changes by less than 0.2
and the derived parameters are the same.
The best model fit to the quiescent light curve and the  different components
in the model i.e. the accretion disk, the  X-ray heated secondary star and the
bulge, are shown in  Figures~\ref{fig:fitq} and \ref{fig:fitqc} respectively.
Changing the secondary star's albedo to 0.3 and 0.5 does not change 
the $\chi^2$ of the fits significantly; the $\chi^2$ changes by only 0.05. 
Also, using an gravity darkening exponent of 0.10 does not change the fits; 
the $\chi^2$ changes by only 0.07.
With the execption of $W$, within the uncertainties, the derived parameters 
are not different.
For $W$=0.3 and 0.5 we obtain 
$\log F_{\rm X,0}$=-13.12 and -12.81~erg~cm$^{-2}$~s$^{-1}$ respectively.
Using the distance of 9.6~kpc
and the X-ray flux obtained from the fits, we obtain an unabsorbed X-ray 
luminosity of 1.2$\times 10^{33}$\,erg\,s$^{-1}$ for XTE~J2123--058 in 
quiescence.

\begin{figure}
\plotone{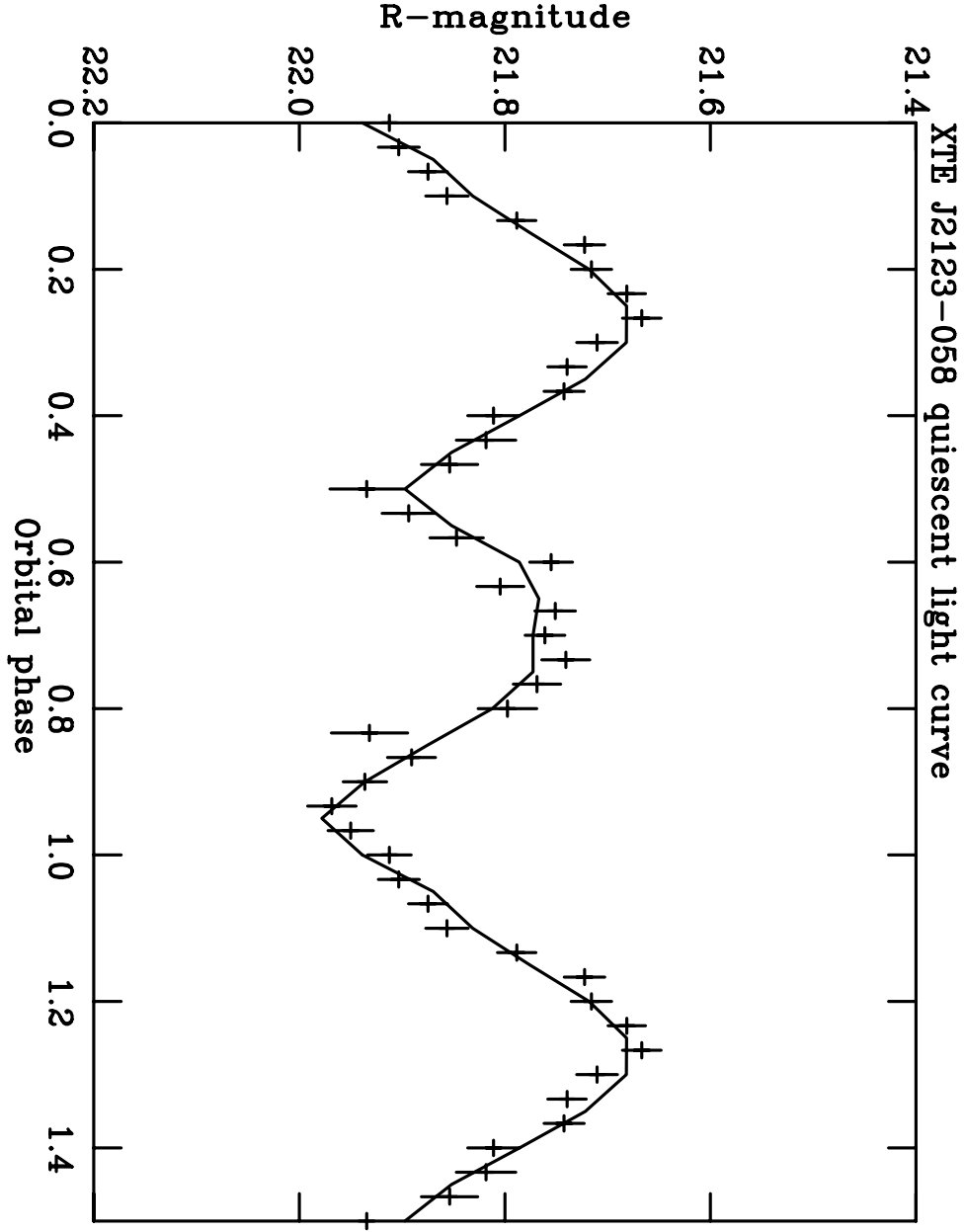}
\caption{The  observed quiescent R-band light curve of XTE~J2123--058 
and a model light curve to show the effects of the accretion disk bulge.
The solid line is a model light curve computed with the X-ray binary model 
with a bulge on the edge of the accretion disk. The bulge has a constant 
height of 17$^{\circ}$ (the flare angle at the edge of the disk) 
and is confined to lie between orbital phase 0.65 and 1.0. 
The effect of the bulge is to eclipse the inner parts of the accretion disk,
thus producing the dip near phase 0.8.
For clarity 1.5 orbital cycles are shown.
\label{fig:bulge} }
\end{figure}

The binary inclination angle we obtain by fitting the quiescent light curve 
is similar to what was obtained by fitting the
outburst data (see section 4.3)
The value we obtain for $\bar{T_{\rm eff}}$ implies a K5 spectral type which is
consistent (within the uncertainties), with the classification of the secondary
star using the absorption lines (see paper II). 
To see if our interpretation of the disk bulge eclipse is correct we
added a bulge to the accretion disk rim. 
To illustrate its effect we compute a model light curve which is shown
in Figure~\ref{fig:bulge}.  We fix the system parameters to the best fit 
values derived earlier. We varied the height of the bulge and the orbital 
phase range where it lies in order to match the observed light curve. 
The bulge was chosen to have a constant arbitrary height of 17$^{\circ}$ 
(i.e. the flare angle at the edge of the disk) and constrained to lie 
between orbital phase 0.65 and 1.0.  The temperature of the bulge is taken 
to be the same as the temperature of the disk's edge.  As one can see, 
the model clearly predicts less flux near phase 0.8, which is due to 
the eclipse of the X-ray heated disk by the cooler disk bulge.

\section{Discussion}

\subsection{The accretion disk radius}
 
We have measurements of the accretion disk radius in XTE~J2123--058 at three
different stages of its outburst; near the peak of outburst, during decay and
in quiescence. The accretion disk radius we measure for XTE~J2123--058 in
quiescence is consistent with that measured in other SXTs. Although
theoretically we cannot comment exactly on how the disk size changes with time,
it should be noted that it is plausible that the disk radius decays
exponentially as is inferred in dwarf novae \citep{Smak84}.

\subsection{The mass of the neutron star}

Observationally, the mean mass of neutron stars in binary radio pulsars is
$1.38 \pm 0.07\,M_{\odot}$ \citep{Thorsett99}. The models of Fryer \& Kaolgera
(2001) agrees well with this observed peak;  they determined the neutron star
and black hole initial mass function, and find that 81-96 percent of neutron
stars lie in the mass range between 1.2--1.6\,$M_{\odot}$. Within this
framework the current mass of the neutron star in XTE~J2123--058
(1.30\,$M_{\odot}$; Table~\ref{tble:mcarlo}) may be the mass at its formation. 
However, in
LMXBs, one expects massive neutrons stars to exist, because the neutron star
has been accreting at the Eddington rate for a long period of time
\citep{Zhang97}.  The observations of kilohertz quasi-periodic oscillations in
some neutron star LMXBs, suggests that neutron stars could be as massive as 
$\sim 2\,M_{\odot}$. If the neutron stars in LMXBs  are formed with a mass of
$\sim 1.4\,M_{\odot}$, then it is quite possible  to accrete an extra 
$\sim 0.6\,M_{\odot}$ in $\sim 10^8$ years \citep{Zhang97}.

\subsection{The secondary star and outburst mechanism}

The  transient  behavior in  LMXBs  has  been diskussed by
\citet{King96}  and \citet{vanParadijs96}. In the model of
\citet{King96}, for a given magnetic braking  law, the mass  transfer
rate is smaller in black hole X-ray binaries  than in neutron star
X-ray  binaries.  Short period   neutron star  X-ray  binaries will
only  be transient if their companions  are highly evolved,  because
the  mass transfer rates in  binaries   with evolved companions are
smaller  than in systems with unevolved   companions. A system will be
transient  if the average mass  transfer rate ($-\dot{M_{2}}$) is
smaller than some critical  value ($\dot{M_{\rm crit}}$).  Although we
find $-\dot{M_{2}}/\dot{M_{\rm crit}}$ to be at least a  factor of 4, obtained 
using our 90 percent mass limits for the binary components and the equations
for $-\dot{M_{2}}$ and $\dot{M_{\rm crit}}$ given in \citet{King96},  
there are many uncertainies in how the mass transfer rates  are determined 
(see paper II); the magnetic braking mechanism is not very well understood
\citep{Kalogera98} and it could be that the mass transfer rate we see
is not the secular mean. Also, XTE~J2123--058  may have had a  very
different evolutionary  path compared  to other SXTs.   The system may
be a  remnant of  a thermal-timescale  mass transfer, leading to
peculiarities close to the end  of the thermal phase 
(King private communication).  

\subsection{The X-ray heated secondary star}

The H$\alpha$ doppler map of XTE~J2123--058 in quiescence presented in 
paper II  shows significant emission arising near the inner Lagrangian 
point, which could
be due to the X-ray heating or chromospheric activity on the secondary
star. Our fits require the inner face of the secondary star to be heated
in order to match the observed shape of the minimum at phase 0.5.  The
dashed line in Figure~\ref{fig:fitq} shows a model light curve 
with no X-ray heating.  It has been known for some time that strong X-ray 
heating in an interacting binary can affect the form of the light curve 
\citep{vanParadijs95}. X-ray heating changes the photometric light curve by
adding light to the minimum at phase 0.5 and changes the radial velocity
curve by shifting the effective centre of the secondary, weighted by the
strength of the absorption lines, from the centre of mass of the star,
resulting in a significant distortion of the radial velocity curve.

The X-ray heating is such that the inner face of the secondary star is
hotter than the outer face.  We can estimate the expected change in
temperature between the inner and outer faces of the secondary star.
Using the X-ray binary model we computed the orbital variation in the
(V-R) color of the secondary star with and without X-ray heating. We
find the color difference between orbital phase 0.0 and 0.5 to be very
small, 0.02 mags, which corresponds to a temperature change of only
$\sim$60~K. Such a small temperature change would be very difficult to
detect.

In order to investigate the effects of X-ray heating we computed the
radial velocity of the secondary star using a crude treatment for X-ray
heating \citep{Phillips99}, where we impose the condition that if the 
incident flux from the X-ray source is more than 50 percent of the intrinsic 
flux from the secondary star then the absorption line flux is set to zero. 
Given the X-ray luminosity we estimate that such a level of irradiation 
will distort the radial velocity curve of the secondary star by 
$\sim$3 percent. However, given the faintness and hence the accuracy 
of the radial velocity curve, this effect would not be currently measurable.

The late-type companion stars in X-ray transients are perfect candidates
to show chromospheric activity. This is because, these stars are tidally
locked in short orbital periods and hence have rapid rotation rates.  
The chromospheric activity would be present in the form of narrow
H$\alpha$ and Ca\textsc{ii} emission. 
Although \citet{Casares01} detect H$\alpha$ emission at the inner
Lagrangian point,  it is difficult to say if this emission is solely due to 
chromospheric activity since the H$\alpha$ emission  can also be 
powered by X-ray heating. However, we can ask how much
of the  H$\alpha$ emission can in principle be powered by X-ray heating.
To do this we need to compare this predicted
emission rate from the X-ray source with the observed H$\alpha$ photon
emission rate.
 
The rate of emission of photons $N_{p}$ with a given energy is given by

\begin{equation}
N_{p}=\frac{F_{p} 4 \pi d^{2}}{E_{p}}
\end{equation}
 
\noindent 
where $F_{\rm p}$ is the photon flux, $d$ is the distance to the source
and $E_{\rm p}$ is the energy of the photon.
From the H$\alpha$ Doppler map (paper II) we estimate the fractional
contribution of the emission on the secondary star relative to the total 
H$\alpha$ emission. We then use this fraction with the 
equivalent width of the whole H$\alpha$ emission line to
estimate the width of the narrow component, which we find to be 
$\sim$1\AA. 
This combined with the dereddened R-band 
magnitude [$m_{R}$=21.8); E$_{B-V}$=0.12 \citep{Hynes01}] 
thus gives an H$\alpha$ flux photon flux of
1.7$\times 10^{-18}$\,erg\,cm$^{-2}$\,s$^{-1}$ 
which corresponds to an emission rate of 
$N_{\rm H\alpha}=4.1\times 10^{40}$ photon~s$^{-1}$ 
($d_{\rm kpc}$=9.6 and $E_{H\alpha}$=1.9eV).

In order to calculate the X-ray photon flux we assume that
each X-ray photon causes one photoionisation of H\textsc{i} and a fraction
$\eta$ then recombine to produce an H$\alpha$ photon.  The
fraction is given by the ratio of the recombination rate of H$\alpha$ to
the total recombination rate of H\textsc{i} as a whole and is taken to be
0.45 for Case B recombination \citep{Hummer87}.  One also has to
allow for the fact that only a fraction of the high energy photons
actually cause photoionisation. 
This fraction $f_\Omega$, depends on the solid
angle subtended by the secondary star and the accretion disk at the compact
object, which is
$\Omega_{\rm star} = \pi(R_2/a)^{2}$ (=0.28) for the star, 
whereas the planar disk
which shields the secondary star subtends a solid angle of 
$\Omega_{\rm disk}=4 (R_d/a)\tan\alpha$ 
(=0.04; where $R_d$ is the accretion disk and 
$a$ is the binary separation). 
Using these values we estimate $f_\Omega$=0.24.
Finally one needs to allow for the X-ray albedo of the secondary star
which is taken to be 0.4 \citep{deJong96}. Assuming that the inferred 
unabsorbed X-ray flux (see section 6.1)  has a mean energy of 2~keV, 
a distance of 9.6~kpc and allowing for factors mentioned above, we obtain 
the the X-ray photon flux of
$N_{\rm X-ray}=1.0\times 10^{41}$ photon~s$^{-1}$.  Since
the predicted number of X-ray photons that produce H$\alpha$ photons is
larger than the actual number observed, we
conclude that the H$\alpha$ observed from/near the secondary star can in
principle by powered by X-ray heating.

\subsection{Comparison with other neutron star SXTs}

The X-ray luminosity in quiescence we derived for XTE~J2123--058 is 
comparable with that observed in other neutron star SXTs, typically 
10$^{32-33}$\,erg\,s$^{-1}$. Although there is some debate as to the 
origin of these quiescent X-rays, accretion seems to be the dominant 
source. \citet{Brown98} suggest that the X-rays could 
be due to cooling of the
neutron star's crust which is heated during outburst. However, as pointed
out by \citet{Narayan01}, neutron star SXTs have 
about half their total 
luminosity in their X-ray power-law tails, which is most
likely to be produced by accretion rather than crustal cooling.  Also,
recent Chandra X-ray observations of Aql X--1 in quiescence show that the
X-ray luminosity varies on timescales which are not consistent with the
heated neutron star crust model \citep{Rutledge01}. It should be
noted that the observed X-ray flux in XTE~J2123--058  is also too high
to be due to the coronal emission from the companion.  
Following  \citet{Bildsten00}, we estimate the X-ray/bolometric flux ratio 
$F_{\rm X,0}/F_{\rm bol}$ to be 3.7, i.e. similar to what is observed 
in other quiescent neutron star SXTs.

Normally the X-ray luminosities of SXTs are obtained by direct X-ray
observations of the source. Our optical study of XTE~J2123--058 in
quiescence has allowed us to infer the quiescent X-ray luminosity. What
would be of great interest is a direct confirmation of the X-ray flux
through direct measurements.

\acknowledgments

TS was supported by a EC Marie Curie Fellowship HP-MF-CT-199900297.
GD is supported by NASA grants NAG 5-7007 and NAG 5-7034.
The United Kingdom Infrared Telescope is operated by the Joint Astronomy
Centre on behalf of the U.K. Particle Physics and Astronomy Research
Council. The WHT is operated on the island of La Palma by the Isaac 
Newton Group in the Spanish Observatorio del Roque de los Muchachos 
of the Instituto de Astrof\'\i{}sica de Canarias.
Based on observations made with ESO Telescopes at the La Silla 
under programme ID 65.H-0509.

\clearpage

\begin{deluxetable}{lccc}
\tabletypesize{\scriptsize}
\tablecaption{Log of observations\label{tble:log} }
\tablewidth{0pt}
\tablehead{
\colhead{Telescope } & 
\colhead{Date} & 
\colhead{Exposure time} &
\colhead{Band }
}
\startdata

WHT 4.2m   &   28 Jun 1999  &	7$\times$530~s  & R  \\
WHT 4.2m   & 	6 Jul 1999  & 157$\times$180~s  & R  \\
WHT 4.2m   & 	7 Jul 1999  &	1$\times$600~s  & I  \\
WHT 4.2m   & 	7 Jul 1999  &	1$\times$600~s  & V  \\
WHT 4.2m   & 	7 Jul 1999  &	1$\times$1200~s & B  \\
WHT 4.2m   &   21 Jul 1999  &  15$\times$600~s  & R  \\
Bok 2.3m   & 4--7 Sep 1999  &  68$\times$1200~s & R  \\
UKIRT 3.5m & 	4 Sep 1999  &  36$\times$60~s   & K  \\
UKIRT 3.5m & 	6 Sep 1999  &  54$\times$60~s   & J  \\
UKIRT 3.5m &   10 Sep 1999  &  54$\times$60~s   & H  \\
NTT 3.5m   &   27 Aug 2000  &  39$\times$600~s  & R  \\
INT 2.5m   & 	5 Nov 2000  &	5$\times$1800~s & U  \\
INT 2.5m   & 	5 Nov 2000  &	1$\times$900~s  & U  \\
INT 2.5m   & 	5 Nov 2000  &	1$\times$600~s  & U  \\
INT 2.5m   & 	5 Nov 2000  &	1$\times$300~s  & U  \\
\enddata
\end{deluxetable}

\begin{deluxetable}{lccc}
\tabletypesize{\scriptsize}
\tablecaption{Observed quiescent average magnitudes of XTE~J2123--058
\label{tble:mags}}
\tablewidth{0pt}
\tablehead{
\colhead{Band  } & 
\colhead{Date} & 
\colhead{Observed magnitude} &
\colhead{reddening$^{*}$ }
}
\startdata
 U         &  5 Nov 2000  & 23.68 $\pm$ 0.02 & 0.56   \\
 B         &  7 Jul 1999  & 23.75 $\pm$ 0.11 & 0.49   \\
 V         &  7 Jul 1999  & 22.65 $\pm$ 0.06 & 0.37   \\
 R         & 21 Jul 1999  & 21.81 $\pm$ 0.02 & 0.27   \\
 I         &  7 Jul 1999  & 21.08 $\pm$ 0.04 & 0.18   \\
 H         & 10 Sep 1999  & 19.48 $\pm$ 0.49 & 0.06   \\
 K         &  4 Sep 1999  & 19.98 $\pm$ 0.48 & 0.04   \\
\enddata
\newline
$^{*}$ $A_{\rm V}/E_{\rm B-V}=3.1$; \citet{Howarth83} and \citet{Seaton79}.
\end{deluxetable}

\begin{deluxetable}{ll}
\tabletypesize{\scriptsize}
\tablecaption{A list of the key variables used in this paper and what they 
represent.\label{tble:defs}}
\tablewidth{0pt}
\tablehead{
\colhead{ } & 
\colhead{ }
}
\startdata

$M_{\rm 1}$   & Mass of the compact object \\
$M_{\rm 2}$   & Mass of the secondary star \\
$q$           & Binary mass ratio defined as $M_{\rm 2}$/$M_{\rm 1}$ \\
$i$           & Binary inclination \\
$d_{\rm kpc}$ & Distance \\
$P_{\rm orb}$ & Orbital period \\
$K_{\rm 2}$   & Secondary star's radial velocity semi-amplitude \\
$\bar{T_{\rm eff}}$ & Mean effective temperature of the seondary \\
$\beta$             & Gravity darkening  exponent \\
$W$            & Albedo of the secondary star \\
$R_{\rm disk}$ & Accretion disk radius \\
$\alpha$       & Accretion disk flare angle \\
$T_{\rm out}$  & Outer disk edge temperature  \\
$\xi$          & Exponent on the power-law radial temperature  distribution  \\
$F_{\rm X,0}$  & Unabsorbed  X-ray flux \\
$A_{\lambda}$ & Interstellar reddening at a given wavelength $\lambda$
\enddata
\end{deluxetable}

\begin{deluxetable}{lccc}
\tabletypesize{\scriptsize}
\tablecaption{Results obtained from the fits to the outburst and decay data
\label{tble:fitod}}
\tablewidth{0pt}
\tablehead{
\colhead{Paramter } & 
\colhead{Best fit} & 
\colhead{68\% confidence} &
\colhead{90\% confidence}
}
\startdata

{\rm Outburst data}    &      &                      &         \\
                       &      &                      &         \\
$q$                    & 0.36 & +0.01~$-$0.01 & +0.16~$-$0.02 \\
$i$  ($^{\circ}$)      & 72.5 & +0.7~$-$0.7   & +1.0~$-$2.3 \\   
$R_{\rm disk}$ ($R_{\rm RL1}$) & 0.69 & +0.01~$-$0.02 & +0.03~$-$0.03 \\
$\alpha$ ($^{\circ}$)  & 5.9  & +0.2~$-$0.5   & +0.5~$-$0.7 \\
                       &      &                      &         \\
{\rm Decay data}       &      &                      &         \\
                       &      &                      &         \\
$R_{\rm disk}$ ($R_{\rm RL1}$) & 0.56 & +0.03~$-$0.03 & +0.05~$-$0.05 \\
$\alpha$  ($^{\circ}$) & 5.2  & +0.5~$-$0.6   & +0.9~$-$0.8 \\ 
\enddata
\end{deluxetable}

\begin{deluxetable}{lccc}
\tabletypesize{\scriptsize}
\tablecaption{Results of the Monte Carlo simulation: the binary masses
\label{tble:mcarlo} }
\tablewidth{0pt}
\tablehead{
\colhead{Paramter } & 
\colhead{Best fit} & 
\colhead{68\% confidence} &
\colhead{90\% confidence}
}
\startdata
$M_{1}$ ($M_{\odot}$) & 1.30 & 1.04~$-$1.70 & 0.95$-$1.92   \\
$M_{2}$ ($M_{\odot}$) & 0.46 & 0.36~$-$0.83 & 0.32~$-$0.98  \\
$R_{2}$ ($M_{\odot}$) & 0.59 & 0.54~$-$0.74 & 0.53~$-$0.76  \\   
\enddata
\end{deluxetable}

\begin{deluxetable}{lccc}
\tabletypesize{\scriptsize}
\tablecaption{Results obtained from the fit to the  quiescent data
\label{tble:fitq} }
\tablewidth{0pt}
\tablehead{
\colhead{Paramter } & 
\colhead{Best value} & 
\colhead{68\% confidence} &
\colhead{90\% confidence}
}
\startdata
$i$                     & 71.8  &  73.0~$-$69.0    & 73.5~$-$~67.0 \\
$R_{\rm disk}$ ($R_{\rm RL1}$) & 0.45 & 0.24~$-$~0.66 &  0.10~$-$~0.80     \\
$\bar{T_{\rm eff}}$ (K) &   4610 &  4575~$-$~4645     &  4560~$-$~4680  \\
$\xi$                   & -1.2  & -1.5~$-$~-0.6    & -1.8~$-$~-0.4 \\
$\log F_{\rm X,0}$ ($\rm erg~cm^{-2}~s^{-1}$)  & -12.99 & -12.92~$-$~-13.08  
& -12.88~$-$~-13.17  \\
$T_{\rm out}$ (K)      &   1425  &   300~$-$~4000     &    $<$~5000  \\ \\
\enddata

\end{deluxetable}

\begin{deluxetable}{lcc}
\tabletypesize{\scriptsize}
\tablecaption{Accretion disk radius evolution \label{tble:rdisk}}
\tablewidth{0pt}
\tablehead{
\colhead{$\Delta t^{*}$ (days) } & 
\colhead{$R_{\rm disk}$ ($R_{\rm L1}$) } & 
\colhead{ $\log F_{\rm X}$ ($erg~cm^{-2}~s^{-1}$)} 
}
\startdata
29   & 0.69$\pm$0.02   &        -7.77      \\
48   & 0.56$\pm$0.03   &        -8.77      \\
551  & 0.45$\pm$0.21   &       -12.99     \\
\enddata
\end{deluxetable}

\clearpage

\end{document}